\RequirePackage{fix-cm}
\documentclass[smallextended]{svjour3}       
\smartqed  
\usepackage[dvipdfmx]{graphicx}
\usepackage[dvipdfmx]{color}
\usepackage{amsmath}
\usepackage{multirow} 
\usepackage{float}
\usepackage{url}
\usepackage{comment}
\usepackage{bm}
\usepackage[T1]{fontenc}

\begin{document}

\title{Visualization for interactively adjusting the de-bias effect of word embedding

}


\author{Arisa Sugino  \and
            Takayuki Itoh  
}


\institute{Arisa Sugino, Takayuki Itoh \at
              2-1-1 Otsuka, Bunkyo-ku, Tokyo 112-8610 Japan \\
              \email{itot@is.ocha.ac.jp}           
}


\maketitle

\begin{abstract}

Word embedding, which converts words into numerical values, is an important natural language processing technique and widely used.  One of the serious problems of word embedding is that the bias will be learned and affect the model if the dataset used for pre-training contains bias.  On the other hand, indiscriminate removal of bias from word embeddings may result in the loss of information, even if the bias is undesirable to us. As a result, a risk of model performance degradation due to bias removal will be another problem.

As a solution to this problem, we focus on gender bias in Japanese and propose an interactive visualization method to adjust the degree of debias for each word category. Specifically, we visualize the accuracy in a category classification task after debiasing, and allow the user to adjust the parameters based on the visualization results, so that the debiasing can be adjusted according to the user's objectives. In addition, considering a trade-off between debiasing and preventing degradation of model performance, and that different people perceive gender bias differently, we developed a mechanism to present multiple choices of debiasing configurations applying an optimization scheme.

This paper presents the results of an experiment in which we removed the gender bias for word embeddings learned from the Japanese version of Wikipedia. We classified words into five categories based on a news corpus, and observed that the degree of influence of debiasing differed greatly among the categories. We then adjusted the degree of debiasing for each category based on the visualization results.  

\end{abstract}


\section{Introduction}

Word embedding is one of the most common techniques in natural language processing. Word embedding is a method of converting words into numerical vectors in a latent space so that semantic relations between words can be handled quantitatively. For example, by expressing words numerically as follows, words with high similarity are learned to be close to each other. Furthermore, when the relationship between words is similar, their difference vectors are also similar.
\begin{eqnarray}
``dog'' = [0.1, 0.2, 0.3, ...] \nonumber \\
``cat'' = [0.2, 0.4, 0.6, ...] \nonumber
\end{eqnarray}

However, word embedding has A problem that the biases inherent in the dataset used to pre-train the language model are directly reflected in the model. When textual data reflecting social values and prejudices are trained, correlations between words containing those prejudices are incorporated into the model.

The main cause of bias stems from the pattern of occurrence of biased words in the text data used in pre-training. For example, if ``woman'' and ``childcare'' often co-occur in the text data, ``childcare'' may have a bias toward ``woman.''  In addition to gender, other attributes, such as religion and race, can also be a factor in bias. The presence of biases in a model may cause problems such as a decrease in the reliability of the model or a lack of fairness.

De-biasing methods that have been studied to solve the bias problem in word embedding include methods that mitigate bias according to specific criteria, such as gender or religion, and re-training using non-biased data. However, all of these methods have the problem of degrading model performance. De-biasing causes a loss of information in the model, making it impossible for the model to accurately understand the meaning and context of words. Bias and debias in word embedding is an issue that is still the subject of active research. Here, most of the previous studies have focused mainly on English embedding.

We believe that what bias refers to and how problematic it is depends on individual values and the situations in which natural language processing models are used.
Based on this premise, we propose a visualization method to achieve debiasing for Japanese nouns on Wikipedia, using gender bias as an example, in which the user can specify the degree to which the bias is mitigated, and furthermore, the degree of debiasing for each word group in a particular category can be adjusted. We propose a visualization method to realize a debiasing for each word group of a specific category. This study aims to suppress performance degradation to the entire model due to debiasing by flexibly setting the degree of debiasing for each category, rather than treating a group of words as a whole.

Since the definition of bias is a subjective issue that varies from person to person and situation to situation, it is not always reasonable to automatically determine the presence or absence of gender bias numerically. Therefore, we introduced visualization so that users can visually understand the status of bias and set the debias configuration appropriately. Specifically, based on a word categorization task, we introduced a method to visualize the debiasing effect for each category and its impact on model performance.
This approach allows the user to visually check the semantic change of a word before and after debiasing and adjust the degree of debiasing, thus enabling word embedding to be debiased according to the intended use. By adjusting parameters for each category, rather than debiasing the model all at once, we aim to achieve debiasing that adaptively suppresses model performance degradation.

\section{Related Work}

\subsection{Bias evaluation for word embedding}

In reducing bias in word embedding, there are many studies on bias measurement methods and a wide variety of approaches. Some studies have surveyed and compared papers on bias analysis \cite{Blodgett20}, others have focused on bias evaluation criteria, and still others have categorically analyzed fairness \cite{Chen24}.
As one of the most famous examples, Bolukbasi et al. \cite{Bolukbasi16} showed that word embeddings can express word analogies by adding and subtracting vectors.
Furthermore, as a result, they showed that word embeddings contain a gender bias, since words that are gender-neutral are associated with one gender. The distribution of gender bias and the strength of the bias were visualized by classifying words into two groups: gender-neutral words and gender-specific words, mainly occupational nouns and words that are associated with gender due to their cultural backgrounds.

As an example of evaluating bias for attributes other than gender, Hutchison et al. \cite{Hutchison20} used BERT (Bidirectional Encoder Representations from Transformers) \cite{Devlin19} to evaluate bias for mental and physical illness. 
Dixon \cite{dixon18} assessed bias related to sexual preference, race, and religion by pre-training ``harmful sentences with bias'' and ``harmless sentences without bias'' for words associated with discrimination.

The above studies only considered bias for words with specific meanings, such as nouns related to occupation or sentences associated with discrimination. On the other hand, this study measures gender bias for all gender-neutral nouns in the dataset without limiting it by word meaning, from the perspective that we should aim for debiasing without limiting its use, assuming that the user is the downstream task of the model.

\subsection{De-bias methods for word embedding}

Sun et al. \cite{sun19} classified debias methods into two categories: data-level methods that remove the bias-causing data from the training data, and model-level methods that suppress the effect of bias by setting constraints that prevent overreliance on specific data.

A representative example of model-level methods is Hard Debias \cite{Bolukbasi16} proposed by Bolukbasi et al. Hard Debias uses subspace projection, and many subsequent debiasing methods \cite{wang20}\cite{liang20}adopt a similar approach.

As for data-level methods, Lu et al. \cite{lu18} proposed a data-level mitigation method for gender bias called Counterfactual Data Augmentation (CDA), which adds gender-reversed data to the trained dataset and does not introduce grammatical inconsistency. Elazar \cite{elazar18} also presented a debiasing method using adversarial learning. However, these studies do not focus primarily on the impact of debiasing on the model.

Some debiasing studies have incorporated visualization techniques, such as deb2viz  \cite{Gyamfi19} by Gyamfi et al and VERB \cite{Rathore24} by Rathore et al.  On the other hand, these studies focused on assistance of users' debias processes, and their main objective was not to analyze the degradation of model performance due to debias.

Regardless of the approach to bias, Sun et al. \cite{sun19} pointed out that debiasing can degrade model performance. Therefore, this study compares the performance of models before and after debiasing and analyzes the impact of debiasing on models.

\subsection{Evaluation of the impact of de-biasing on the model}

All of the aforementioned debiasing methods of word embedding have the problem of degrading the performance of the model. For this reason, the relationship between model performance degradation and debias is still a subject of research. Kobayashi et al. \cite{kobayashi22} showed that ``increased representation ambiguity'' in word embedding can cause model performance degradation.  Also, this study offered a visual analytics system for evaluation the debiasing effect and that it allows the user to analyse the missclassified news articles and reason of misclassification.
Ilzu \cite{iluz24} also evaluated the effect of debiasing on machine translation using an automatic accuracy metric \cite{Stanovsky19}.

The above studies use the existing debiasing method \cite{Bolukbasi16} and do not propose improvements to the debiasing method itself. In addition, all of these studies comprehensively treat the entire model, and there are no visualization or debiasing methods that focus on specific words. In this study, we aim to improve the debiasing by manipulating the parameters of the debiasing method for each word category while using visualizations, so that the impact on the model is reduced.

\subsection{Visualization for gender bias}

Visualization of data bias has been a well-discussed issue, especially focusing on the fairness of machine learning.
Ahn et al. \cite{Ahn19} presented FairSight that visualizes the three phases, data processing (Data), selection of learning models (Model), and generation of rankings as learning outcomes (Outcome), to achieve the visualization focusing on fairness.
Cabrera et al. \cite{Cabrera19} presented FairVis, a visual analytics system to prevent discrimination and unequal learning outcomes by grouping compositely sensitive attributes such as race and gender.
Wang et al. \cite{Wang20b} presented DiscriLens which supports comprehensive visual analytics of bias in machine learning by combining a discrimination detection module with a visualization module.
Tochigi et al.\cite{Tochigi21} visualized the bias of machine learning in the recommendation system by displaying the difference between the viewing history and the recommendation results in the machine learning movie recommendation system data.

Visualization of gender differences has also been an active topic.
Tovanich et al. \cite{Tovanich21} presented the statistical difference of publications and careers of male and female researchers working on computer visualization for 30 years.
Cabric et al. \cite{Cabric23} presented the representation change of gender differences in visualization research.
Nakai et al. \cite{Nakai23} developed a hierarchical data visualization technique specific to the representation of gender differences.

\section{Debiasing process and visualization for its adjustment}

\subsection{Processing flow}

This study first performs a gender bias mitigation process for word embeddings. Using the debiased word vectors, the HNSW (Hierarchical Navigable Small World) algorithm \cite{Malkoy16} is used to search for the vector with the highest similarity among the word vectors before debiasing. The text data of the words associated with the resulting vectors is obtained. Based on the text data of the words, a category classification task using BERT (Bidirectional Encoder Representations from Transformers)\cite{Devlin19} is performed. The changes in category labels before and after debiasing are compared and analyzed through visualization to evaluate the degree to which the word vector has changed semantically as a result of the debiasing process. This process makes it possible to quantitatively understand the effect of debiasing and its impact.

In this method, we set the degree of debiasing differently for each word category. This allows appropriate adjustment of the bias for each category and improves the fairness of the word vectors.

\subsection{Debias}

\subsubsection{Data structure}

The word embedding data used in this method is defined as a set $A = \{a_1, a_2, \dots , a_n\}$ consisting of $n$ of $m$-dimensional vectors. Each vector $a_i = (x_{i1}, x_{i2}, \dots ,x_{im})$ has a corresponding word label.

\subsubsection{Improvement of Hard Debias focusing on projection vectors}

Our method improves on an existing debias method, Hard Debias \cite{Bolukbasi16}, by introducing a mechanism that allows the degree of debias, described as $\theta$, to be adjusted for each specific word group, thereby mitigating gender bias in word embeddings.  This section first explains the process of Hard Debias and explain how our method improves on it, and then describes the advantages of our method over existing methods.

Hard Debias is a method for removing bias components based on a specific axis in the word vector space, which is determined by the bias to be debiased. For example, to debias gender bias, the axes are created using words that have gender in their meanings, such as ``man'' and ``mother''.

In word embedding, words are converted into a multidimensional numeric vector, and semantic relations between words are expressed in terms of positional relations in the vector space. However, if this vector space contains gender bias, undesirable relations (e.g., ``male = work'', ``female = housework'') may be established at the same time, as described below. Hard Debias is a method that uses vector operations to remove such bias components and mitigate biased relations between words.
$$\overrightarrow{queen} \approx \overrightarrow{king}-\overrightarrow{man}+\overrightarrow{woman}$$
$$\overrightarrow{homemaker} \approx \overrightarrow{computer~programmer}-\overrightarrow{man}+\overrightarrow{woman}$$
In the following, we take the case of mitigating gender bias as an example to explain the processing procedure of Hard Debias. Note that words directly related to gender (e.g., ``man'', ``woman'') are not subject to debias.

\subsubsection*{Definition of gender axis}
First, we define the gender axis using gender-related word pairs. Gender-related word pairs are composed of pairs of words that are paired for men and women, such as ``man''-``woman'' and ``father''-``mother''. Let $P_i$ denote the $i$-th gender word pair consisting of the vector $(vm_i,vw_i)$. For each pair, calculate the vector $c_i$ of midpoints using the following formula.
$$
c_i=\frac{vm_i+vw_i}{|P_i|}
$$
where $c_i$ represents the average vector of the pair $P_i$. Subsequently, Then, the method performs centralization by subtracting $c_i$ from each word vector.
$$
vm_i'=vm_i-c_i
\quad
vw_i'=vw_i-c_i
$$
The method orders the centered vectors from all pairs $P_i$ to generate a matrix $M$, where the size of $M$ is (the number of all words in the pair) $\times$ (the number of dimensions of the word embedding).
$$
M=
\begin{bmatrix}
   vm_1'  \\
   vw_1' \\
   vm_2' \\
   vw_2'\\
   \vdots
\end{bmatrix}
$$
Next, the method obtains the principal components of the group of centralized vectors by computing the covariance matrix using $M$ as follows:
$$
Gender Direction: G=\sum_i\frac{M^\mathrm{T}M}{|Pi|}
$$
By decomposing this covariance matrix into eigenvalues, the method extracts principal components representing the gender bias in the vector space. This principal component vector $g$ is regarded as the gender axis $G$.

\subsubsection*{Centralization}
Next, the gender-neutral word $V$ (e.g., ``doctor'', ``computer programmer'') represented by the vector $V$ to be debiased is orthonormalized to the gender axis $G$. This process removes the gender information from $V$ and reduces the gender bias. First, the method computes the projection vector $proj_G(V)$ of $V$ onto $G$ as follows:
$$
proj_G(V)=\frac{v\cdot g}{|g|^2}g
$$
Finally, the methods computes the debiased vector $v^{debias}$ by subtracting $proj_G(V)$ from $v$.
$$
v^{debias}=v-proj_G(V)
$$

\subsubsection*{Parallelization}
Our method re-aligns the gender-associated words to equidistant and contrasting positions relative to the bias axis. This is an operation to strengthen the equivalence of pairs of gender-related words ($vm_i$,$vw_i$) and to prevent the words after debiasing from being biased toward particular gender words. Let $(vm_i'',vw_i'')$ be the vector after the parallelization, and the following equation is used.
$$
vm_i''=x+y\cdot g \quad
vw_i''=x-y\cdot g
$$$$
x=c_i-proj_G(c_i)\quad
y=\sqrt{1-||x||^2}
$$
where $c_i$ is the midpoint of the pair ($vm_i$,$vw_i$) and $g$ is a vector indicating the gender axis. This parallelization ensures that the post-debiased word vectors are properly contrasted and consistent with the debiasing.

Hard Debias is one of the most important methods in word embedding debias and is widely used as a basis for debias research \cite{wang20}\cite{liang20}. However, Hard Debias completely removes a specific direction in the vector space, so it may happen that the original meaning of a word is not fully preserved. As a result, the performance of downstream tasks such as text classification may be degraded by Hard Debias \cite{sun19}.

Therefore, this study aims to achieve both desirable model performance and sufficient debiasing effect by making the ratio of vector components to be removed during debiasing adjustable. The method introduces a parameter, the ratio of projection vectors subtracted from the word vectors to be debiased, which allows the degree of debiasing to be freely changed.

\subsubsection*{Parameterization of Hard Debias}

While debiasing gender-neutral words $V$, this method uses the size of the projection vector $proj_G(V)$ as the strength of the bias of $V$, and the degree of bias removal can be adjusted arbitrarily by setting the variable $\theta$.
$$
v^{debias}=v-\theta \cdot proj_G(V)
$$
The variable $\theta \in [0,1]$ is a parameter that represents the strength of the debiasing:
\begin{itemize}
  \item $\theta=0$: No Debias
  \item $\theta=1$: Hard Debias
\end{itemize}

This parameterization allows for the control of word meaning loss and provides more flexibility than conventional methods.  We believe that the criteria for gender bias differ from person to person. The ultimate goal of this research is to break away from a fixed definition of gender bias through this improvement, and to achieve fairness that reflects user intentions.

\subsection{Word Categorization}

\subsubsection{BERT-based categorization model using HNSW}

In order to verify the semantic change of words after debiasing, we built a word category classification model trained based on BERT (Bidirectional Encoder Representations from Transformers).

BERT is a natural language processing model developed by Google.  BERT is a model that captures the meaning of words by simultaneously considering the context before and after the text, and is able to utilize the information of the entire sentence better than conventional unidirectional models. Compared to conventional unidirectional models, BERT is more effective in situations that require accurate word meaning understanding, such as classification tasks, because it can utilize information from the entire sentence.

Text classification using BERT usually involves tokenizing and quantifying raw text data to convert the text into a form that can be understood by the model. In this method, we created a BERT-based categorization model using word vectors as input values. The model searches for approximate nearest neighbors using HNSW (Hierarchical Navigable Small World) on the input vectors, and outputs the word category with the most similar vector as the classification result. By using the pre-debiased categorization results as correct data and comparing them to the categorization results using the post-debiased vectors, it is possible to evaluate the change in word meaning due to debiasing.

HNSW is an algorithm for highly accurate approximate nearest neighbor search in high-dimensional spaces. Nearest neighbor search is an optimization problem to find the nearest point from a certain point, and the computational cost increases exponentially when searching from a large amount of data. However, HNSW uses a hierarchical structure that enables fast search even on large data sets. In this method, the HNSW index was created using the word vectors before debiasing to compare how the meanings of words changed before and after debiasing, using the word vectors after debiasing as input values.

\subsubsection{Fine tuning}

Our method applies fine-tuning to BERT using text data and corresponding category labels. Specifically, we used labeled data created from news articles and word-by-word labeled data created by making ChatGPT output a set of words that are highly relevant to each category.

First, we created training data by labeling categories based on article headlines using a news corpus. Using this training data, we fine-tuned the BERT model for the word categorization task. Next, we perfornmed a categorization task on the test data created using GPT-4o, and the final model was retrained on the manually re-labeled training data based on the output results. We created the test data by making GPT-4o enumerate the Japanese nouns associated with each category, and then extracting and labeling the words from the output that were included in the word data to be debiased and that were judged to be appropriate for categorization by direct visual inspection. The test data were the words that were extracted from the output results and labeled as test data.

We determined the number of epochs of fine tuning by comparing Training Loss (loss on training data) and Validation Loss (loss on validation data).  Both Training Loss and Validation Loss decrease in the case of normal training.  On the other hand, Training Loss decreases while Validation Loss begins to increase in the case of over-training.

We evaluated the model performance using Accuracy and F1 Score, where Accuracy is the percentage of correct categorizations and F1 Score is the harmonic mean of the fit and recall. The goodness of fit indicates the proportion of correct predictions that are actually correct, while the recall indicates the proportion of actual correct predictions that the model was able to predict as correct.
By simultaneously checking Accuracy and F1 Score, we can accurately measure the model performance even when the number of samples across classes differs significantly. For example, if 90\% of the entire word vector dataset is distributed in the sports category, then just predicting sports at all times will result in a 90\% correct prediction rate. However, in this case, categories other than sports are not predicted at all.

When the Accuracy and F1 Score are high simultaneously, the model is accurate overall, indicating balanced performance for each class. This method calculates a weighted F1 Score in order to account for the bias in the number of data among classes. The weighted F1 Score enables appropriate evaluation even for classes with a small number of data and more accurately reflects model performance.

\subsection{Adjustment of the degree of debiasing through optimization problems}

Our method computes the Pareto-optimal solution by applying a multi-objective optimization, taking into account the trade-off between the suppression of model performance degradation and the debiasing. Specifically, the Pareto front is obtained based on a problem setting that maximizes model performance (Accuracy and F1 Score) and minimizes gender bias at the same time.

Pareto-optimal refers to a state in which no further improvements can be made to all objectives at the same time, since improving one solution will worsen at least another objective (model performance and de-bias in this method). By enumerating these solutions, this study allows the user to flexibly select the degree of debiasing as a compromise depending on the user's requirements, for example, when priority is given to debiasing or when the degradation of model performance is to be minimized. Furthermore, based on the results of calculating the weighted sum of the model performance and bias weights, we identified the degree of debiasing that has the same level of impact on both. This mechanism is useful since it can concretely present the degree of debiasing in accordance with the user's intention and target of use, even in situations where tradeoffs exist.

As an indicator of gender bias, we calculated the difference between the cosine similarity of words in each category that have a large bias and words that represent males, and the difference between the cosine similarity of words in each category that have a large bias and words that represent females. The closer this value is to 0, the more equidistant the words are from both males and females in the vector space, i.e., the more neutral and gender bias is suppressed.

\subsection{Visualization for adjusting the degree of debiasing}

The method assists users who adjust the degree of debiasing while verifying model performance due to debiasing through the following visualization.
\begin{itemize}
  \item Heatmap of confusion matrix indicating accuracy of categorization. See Fig. \ref{p4:02}, \ref{p4:03}, \ref{p4:04}, and \ref{p4:06}. 
  \item Line chart representing the change in categorization performance with respect to the change in the degree of subtraction $\theta$ of the gender information during debiasing. See Fig. \ref{p4:05}.
\end{itemize}
These specific use cases are described below in the next section.

\section{Experiment}

\subsection{Sample data}
\label{sec_4_2}

We implemented the presented method with Python 3.10 on a MacBook Air (macOS Ventura ver. 13.4). The GPU at Google Colaboratory was used for fine tuning and learning the categorization model.

Our experiment used ``Japanese Wikipedia Entity Vector'' \footnote{\url{https://www.cl.ecei.tohoku.ac.jp/msuzuki/jawiki_vector/}} published by Tohoku University.
This is a Word2vec model \cite{Mikolov13,Mikolov13b} \footnote{{url{https://code.google.com/archive/p/word2vec/}}} learned from the entire text of the Japanese Wikipedia 
In order to debias only Japanese nouns, we used MeCab\footnote{\url{https://taku910.github.io/mecab/#download}} for morphological analysis to exclude words other than nouns.
Furthermore, using regular expressions, only hiragana, katakana, kanji, and long vowels were recognized as nouns, and symbols and alphabetic characters were excluded.

In this study, we created a set of pairs of gender words based on ``Gender Pair'' \cite{lu18} by Lu et al. We first translated all words in ``Gender Pair'' into Japanese to minimize the number of tokens. Cases in which gender was not included in the translation (e.g. ``postwoman'' to ``postal worker'' in Japanese) or not included in the vocabulary of the Wikipedia entity vector dataset used were excluded. These gender words were also excluded from the debiasing.

\subsection{Categorization of bias}
\label{sec_4_3}
\noindent

We aimed to achieve debiasing with minimal degradation of model performance by classifying vectors and biases and adjusting parameters for each word feature, rather than debiasing the model all at once in this study.   We tried the following feature classification methods to archive this goal.  

\subsubsection{Bias clustering by k-means method}
First, we applied the the k-means clusterig to word vectors, and then applied optimal debiasing to each cluster in an attempt to prevent performance degradation of the model after debiasing.  We first used the Elbow method to detect the optimal number of clusters.
The results of the elbow method showed that the line chart was generally smooth even when the number of clusters was increased, and the k-means clustering method was judged to be unsuitable for the dataset in this study. Although there have been examples of successful bias classification by clustering in a previous study, we consider that the clustering of word vectors was achieved while maintaining lexical consistency because this previous study was conducted only for words in a specific field of legal documents. On the other hand, in this study, since the words were not limited based on their topics, it might happen a lack of lexical consistency in the dataset, and clustering by the k-means method was difficult. Therefore we needed to explore word classification methods using approaches other than vector clustering.

\subsubsection{Bias classification by word category classification task}

In this study, we classified words into five categories: entertainment, science, politics, sports, and business.
As a model for BERT, we used a pre-trained Japanese model \footnote{\url{https://github.com/cl-tohoku/bert-japanese}} published by Tohoku University.
This model is based on the Japanese Wikipedia \footnote{\url{https://ja.wikipedia.org/wiki/\%E3\%83\%A1\%E3\%82\%A4\%E3\%83\%B3\%E3\%83\%9A\%E3\%83\%BC\%E3\%82\%B8}}
for the corpus.

We fine-tuned the model for the word categorization task on training data created by labeling categories based on article headlines of livedoor news corpus \footnote{\url{https://www.rondhuit.com/download.html#ldcc}}.   Then, we performed the categorization task on test data created using GPT-4o, and retrained the model on new training data that was manually re-labeled based on the output results.

\begin{figure}[ht]
\centering
\includegraphics[keepaspectratio,width=.8\linewidth]{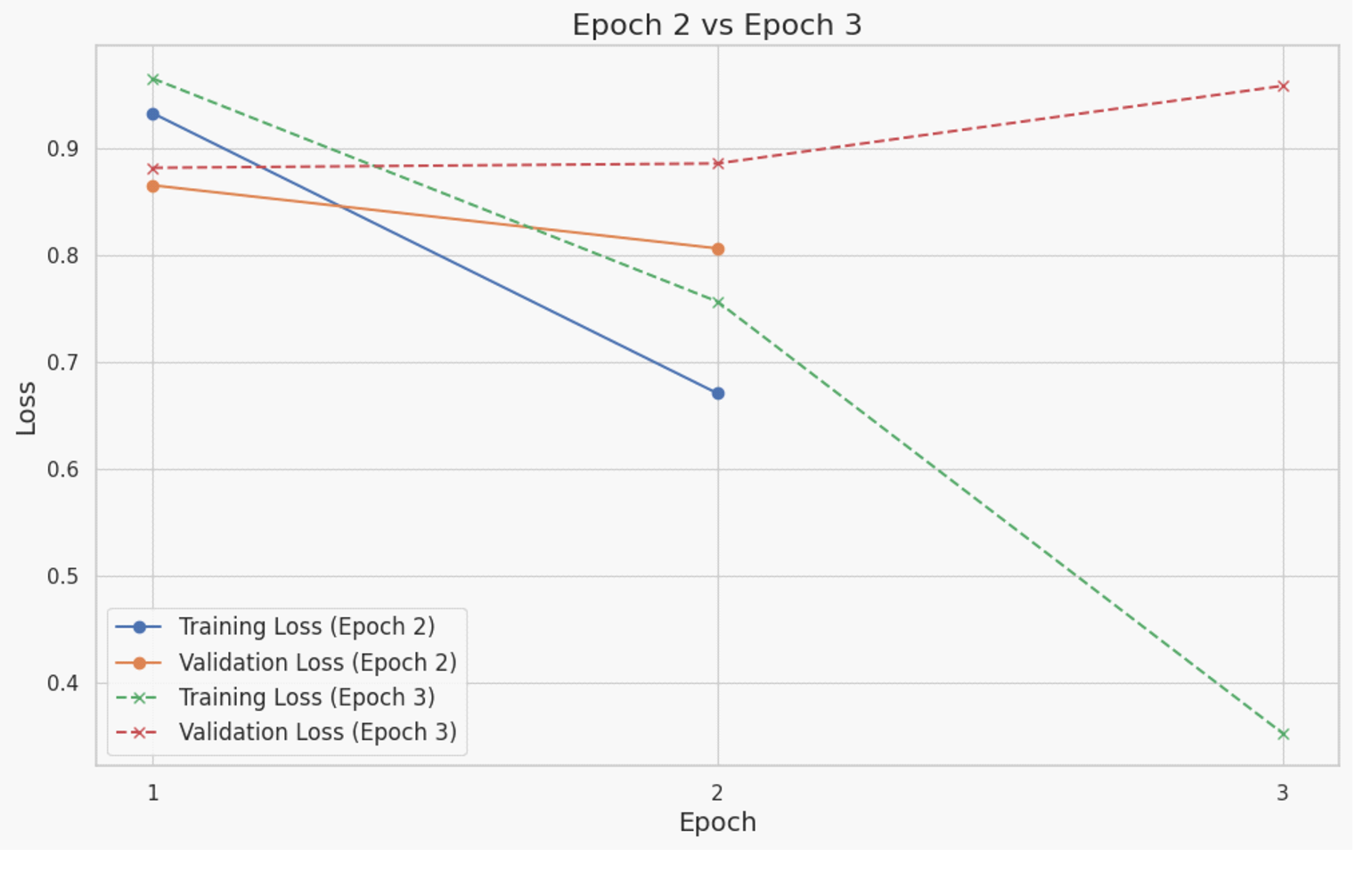}
\caption{Comparison of epoch number and training loss and validation loss.}
\label{p4:loss}
\end{figure}

Figure \ref{p4:loss} shows the training loss and validation loss by number of epochs (Epoch) for the fine-tuning of the trained model for the word classification task.
The horizontal axis plots Epoch, and the vertical axis plots training loss and validation loss.
The solid line indicates Epoch=2, and the dotted line indicates the number of Epoch=3. Since the validation loss at Epoch=3 (red dotted line) showed signs of overlearning, fine tuning was conducted at Epoch=2.

\subsection{Change of word category before and after debiasing}
\label{sec_4_4}
\noindent
\subsection{Category distribution before debiasing}

\begin{figure}[ht]
\centering
\includegraphics[keepaspectratio,width=.8\linewidth]{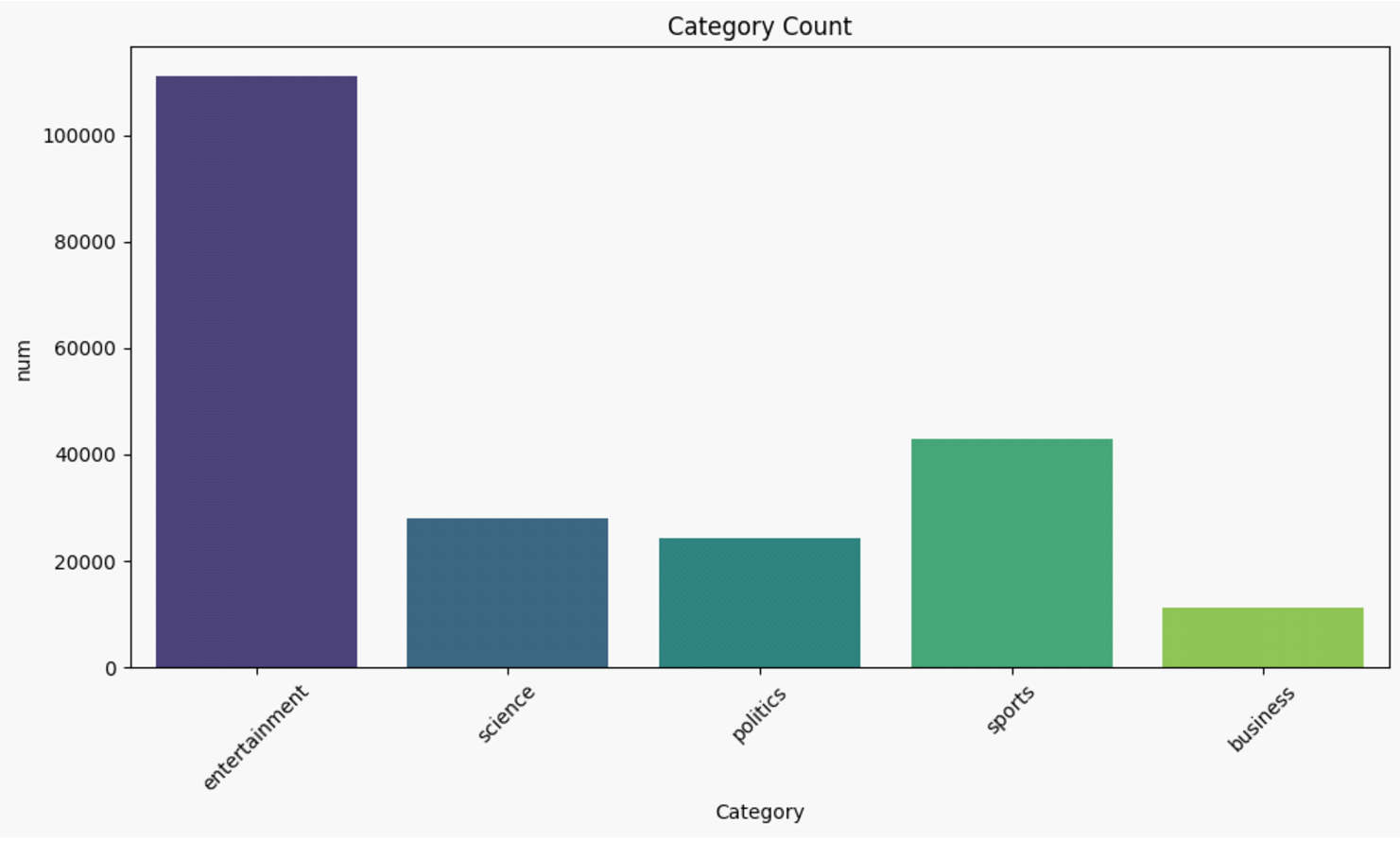}
\caption{Category distribution before debiasing.}
\label{p4:01}
\end{figure}

\begin{figure}[ht]
\centering
\includegraphics[keepaspectratio,width=.8\linewidth]{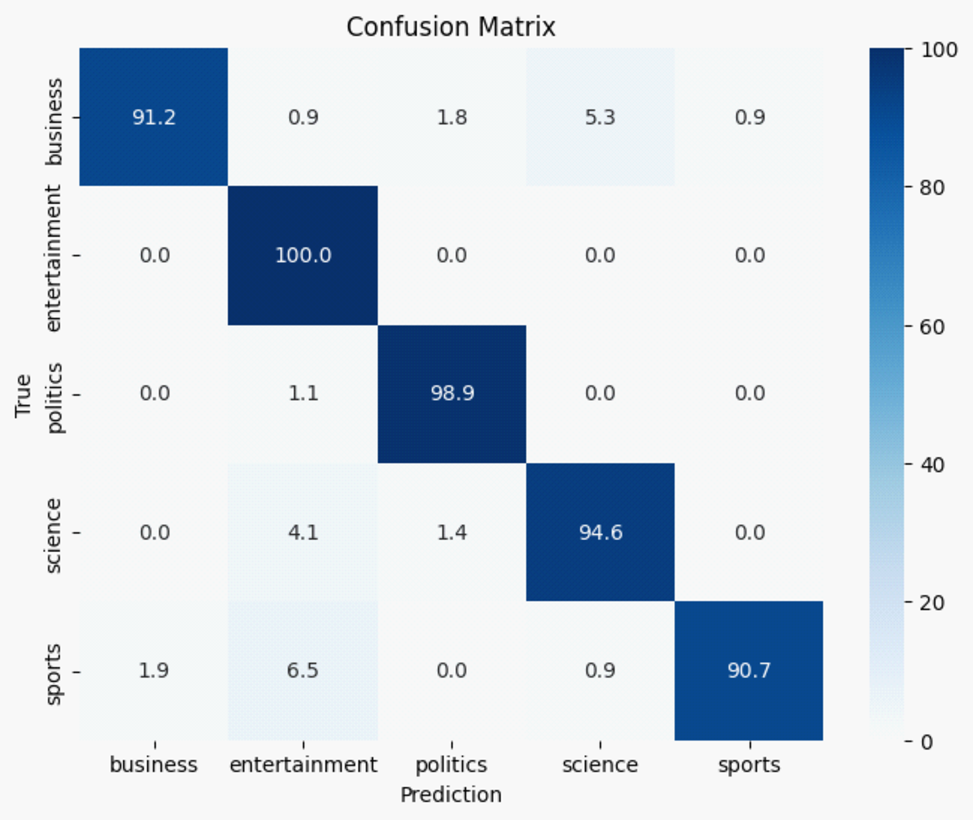}
\caption{Category classification result before debiasing.}
\label{p4:02}
\end{figure}

Figure \ref{p4:01} shows the results of the category classification task without debiasing, where the horizontal axis shows the categories of words while the vertical axis shows the number of words in each category.
Figure \ref{p4:02} shows a confusion matrix that visualizes the accuracy of the classification results. The vertical axis indicates the correct answer labels, while the horizontal axis indicates the predicted labels. The values in the matrix indicate the percentage of the correct labels in each category that are predicted to be in that category. As an example, the 0.9 value in the top right corner indicates that 0.9 percent of the words in the category ``business'' were misclassified to ``sports'' out of the words in the category ``business''. The higher the value of the diagonal line, the more correctly classified the word is.

The high value of the diagonal component of the confusion matrix in Figure \ref{p4:02} confirms that the categories are classified correctly overall. From this and Figure \ref{p4:01}, we found that the pre-debiased words are mostly related to entertainment, with sports, science, politics, and business having the highest number of words, in that order.

\subsubsection{Category distribution after hard debiasing}

\begin{figure}[ht]
\centering
\includegraphics[keepaspectratio,width=.8\linewidth]{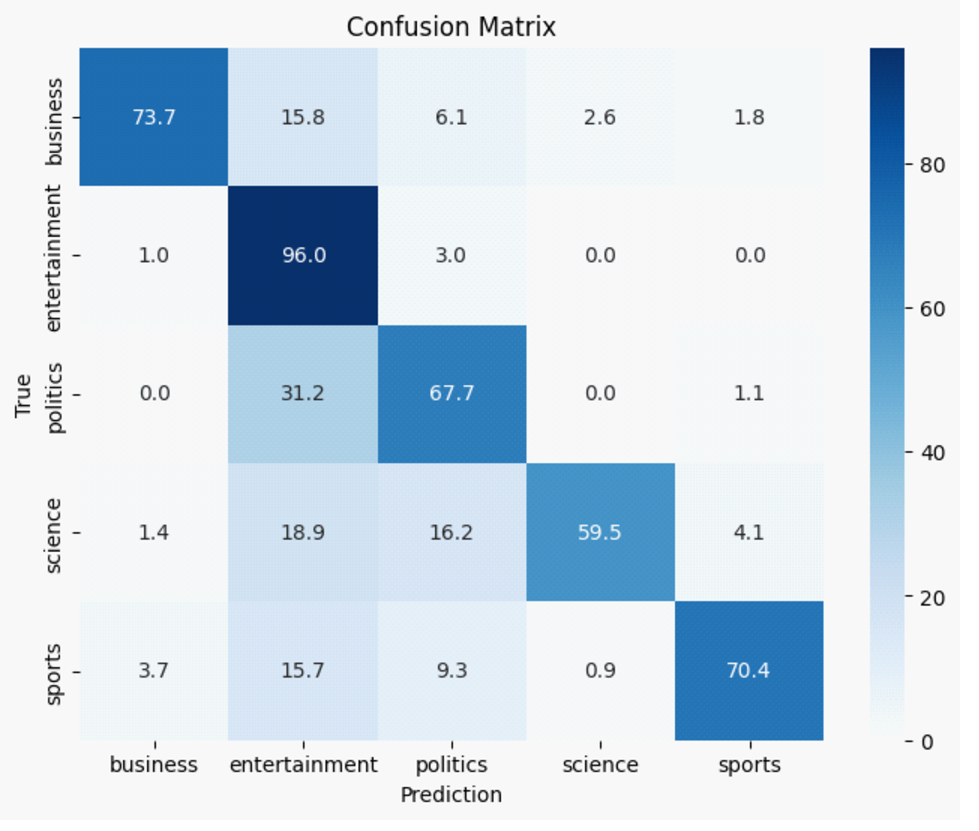}
\caption{Category classification result after hard-debiasing.}
\label{p4:03}
\end{figure}

\begin{figure}[ht]
\centering
\includegraphics[keepaspectratio,width=.8\linewidth]{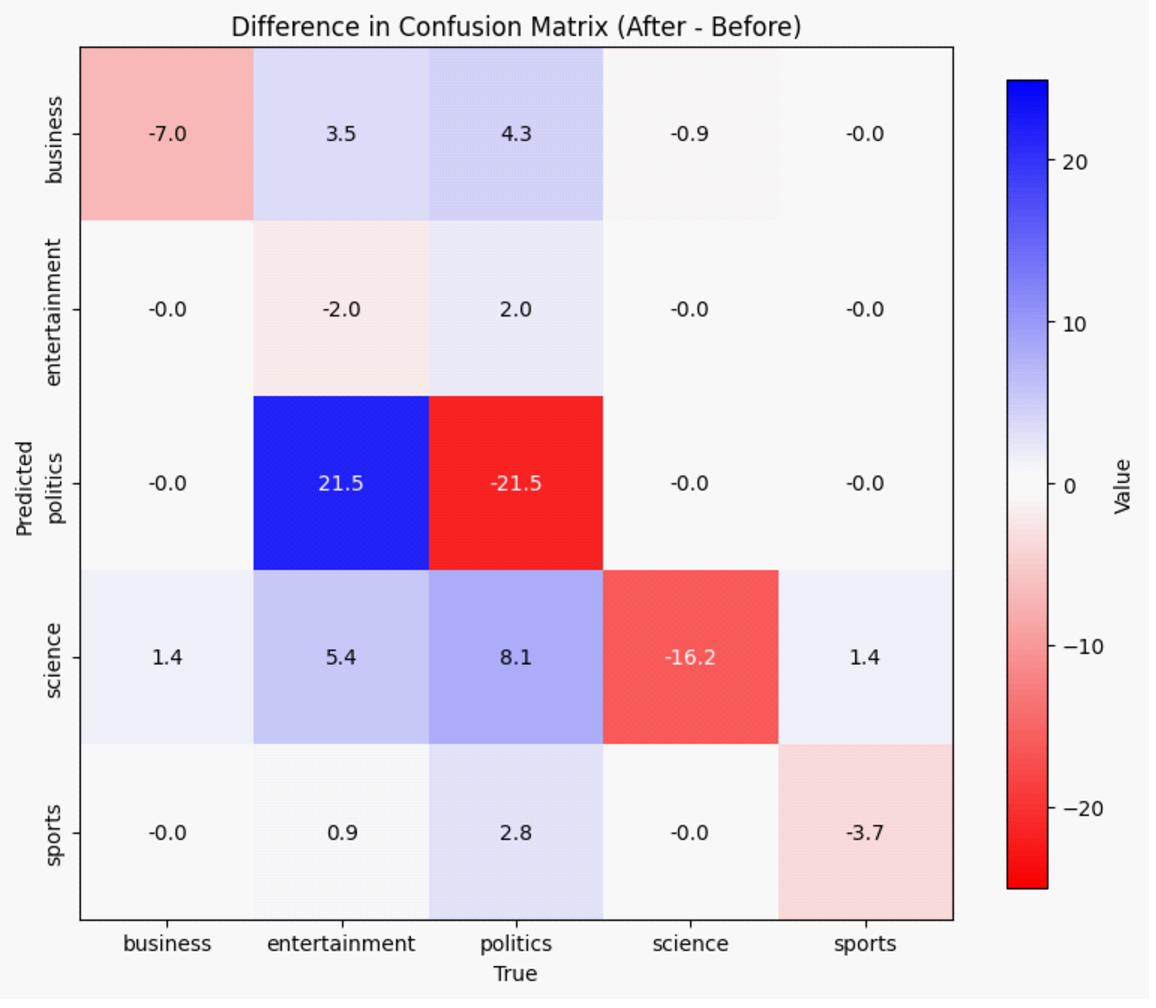}
\caption{Difference in confusion matrices before and after debiasing.}
\label{p4:04}
\end{figure}

Figure \ref{p4:03} shows the confusion matrix representing the categorization results for words that were hard-debiased for all categories.
To confirm the change in categorization before and after debiasing, we visualized the difference in distribution proportions in each category after and before debiasing with a confusion matrix, as shown in Figure \ref{p4:04}.
The vertical axis in Figures \ref{p4:03} and \ref{p4:04} plots the correct answer labels, while the horizontal axis plots the predicted labels. Figure \ref{p4:04} shows the result of subtracting the pre-debiased confusion matrix from the post-debiased confusion matrix.

Focusing on the diagonal component, we found that the debiasing process causes misclassification in all categories. In particular, the accuracy of categorization in the politics and science categories is reduced. Specifically, words in the politics category were misclassified into the entertainment category, and words in the science category were misclassified into either the politics or entertainment category at a high rate. On the other hand, words in the entertainment and sports categories were more likely to be classified into the same category after debiasing.

Here, accuracy and weighted F1 Score were compared before and after debiasing to check for changes in the overall accuracy of the model.
Table \ref{table:4_01} shows that Accuracy and F1 Score decreased after debiasing. This indicates that word embedding was degraded by the debiasing process. The fact that Accuracy and F1 Score decreased to about the same degree confirms that the effect of debiasing was not biased toward any particular category, but appeared evenly.

\begin{table}[hbtp]
  \caption{Accuracy and F1 Score before and after debiasing.}
  \label{table:4_01}
  \centering
  \begin{tabular}{lcr}
    \hline
      & Accuracy  &  F1 Score  \\
      &&(weighted)\\
    \hline \hline
    Before debiasing  & 0.94888  & 0.94890 \\
    After debiasing  & 0.85481   & 0.85743 \\
    \hline
  \end{tabular}
\end{table}


\subsubsection{Debias of specific categories}

We reduced the degree of debiasing for politics, which had the largest impact on debiasing as shown in Figure \ref{p4:04}, and then debiased it again.

\begin{figure}[ht]
\centering
\includegraphics[keepaspectratio,width=.8\linewidth]{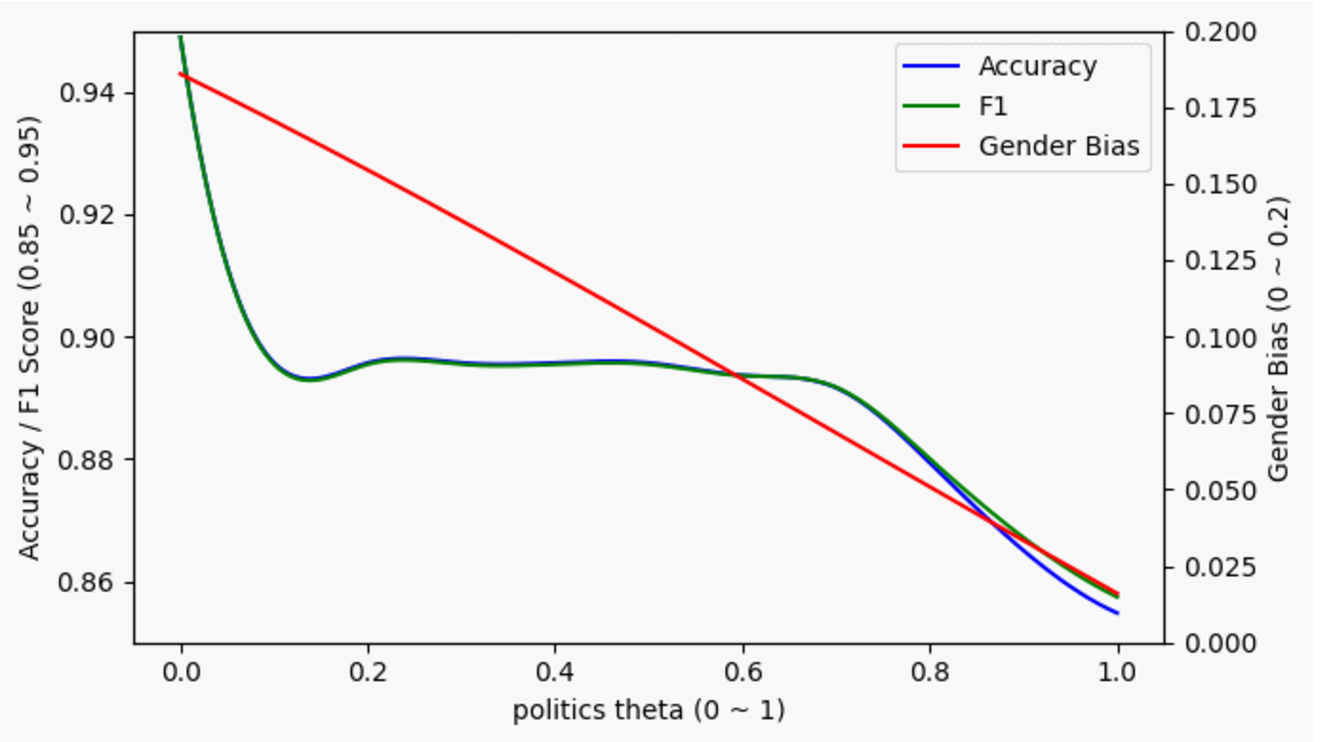}
\caption{Accuracy, F1 Score and bias with the change in debias degree of politics.}
\label{p4:05}
\end{figure}

The degree of subtraction $\theta$ of gender information during debiasing was varied from 0.0 to 1.0 only for the politics category, while $\theta$=1.0 (perfect debiasing) was applied to the other four categories. 
Figure \ref{p4:05} shows the result, where the horizontal axis plots the degree of politics debiasing $\theta$, the left side of the vertical axis plots Accuracy and F1 Score, and the right side of the vertical axis plots the following quantities:

\begin{itemize}
\item Cosine similarity between ``a group of words in politics with a large bias'' and ``a word describing a man.''
\item Cosine similarity between ``a group of words in politics with a large bias'' and ``a word describing a woman.''
\end{itemize}

\noindent
The difference between these two values is taken on the right side of the vertical axis.  Here, the closer to 0 means that the word with the larger bias is more neutralized from gender by the debiasing operation.

As shown in Figure \ref{p4:05}, we can observe a clear tradeoff between the suppression of model performance degradation and debiasing, and it is difficult to uniquely determine the $\theta$ that maximizes both at the same time. However, when $\theta$ was varied from 0.1 to 0.7, Accuracy and F1 Score remained almost flat. Furthermore, when the Pareto front was calculated from the viewpoint of Pareto Optimality, the values of $\theta$ that constitute the Pareto front in Figure \ref{p4:05} were found to be 0.0, 0.6, 0.7, 0.8, 0.9, and 1.0. In addition, we calculated the weighted sum of the model performance and bias with equal weights, and found that when $\theta$=0.7, the degradation of model performance and the effect of bias are equal. In other words, between 0.8 and 1.0 of $\theta$ is suitable when debiasing is more important, and $\theta$=0.0 or 0.6 is desirable when the degradation of model performance is to be suppressed as much as possible.

We calculated $\theta$ by repeating the above steps for each category for which the model performance degradation and the bias effect are comparable.
Table \ref{table:4_02} compares the results for the ``emphasis on debiasing'', the ``emphasis on minimizing performance degradation'', and the ``emphasis on both equally''.

\begin{table}[hbtp]
  \caption{$\theta$ for each category by debiasing condition.}
  \label{table:4_02}
  \centering
  \begin{tabular}{lllll}
    \hline
      & emphasis on minimizing performance degradation  &  both & emphasis on debiasing  \\
    \hline \hline
    politics  & 0.0, 0.6  & 0.7 & 0.8$\sim$1.0\\
    science  & 0.5$\sim$0.7   & 0.8 & 0.9$\sim$1.0 \\
     business  & 0.6   & 0.7 & 0.8$\sim$1.0 \\
    sports  & 0.6   & 0.9 & 1.0 \\
    entertainment  & 0.7$\sim$0.8 &0.9  & 1.0 \\
    \hline
  \end{tabular}
\end{table}

By implementing debiasing using the $\theta$ value corresponding to ``both'' in Table \ref{table:4_02}, the effect of debiasing and the performance degradation of the model can be minimized to the same extent. Within the range of $\theta$ shown in Table \ref{table:4_02}, setting a value larger than the $\theta$ value will place more emphasis on debiasing, while setting a smaller value will allow debiasing that prioritizes maintaining model performance.

Here, we visualized the difference between the debias calculated from Table \ref{table:4_02} and the impact of the existing Hard Debias method on the model, where the impact of debias and the degradation of model performance are kept at the same level.

\begin{figure}[ht]
\centering
\includegraphics[keepaspectratio,width=.8\linewidth]{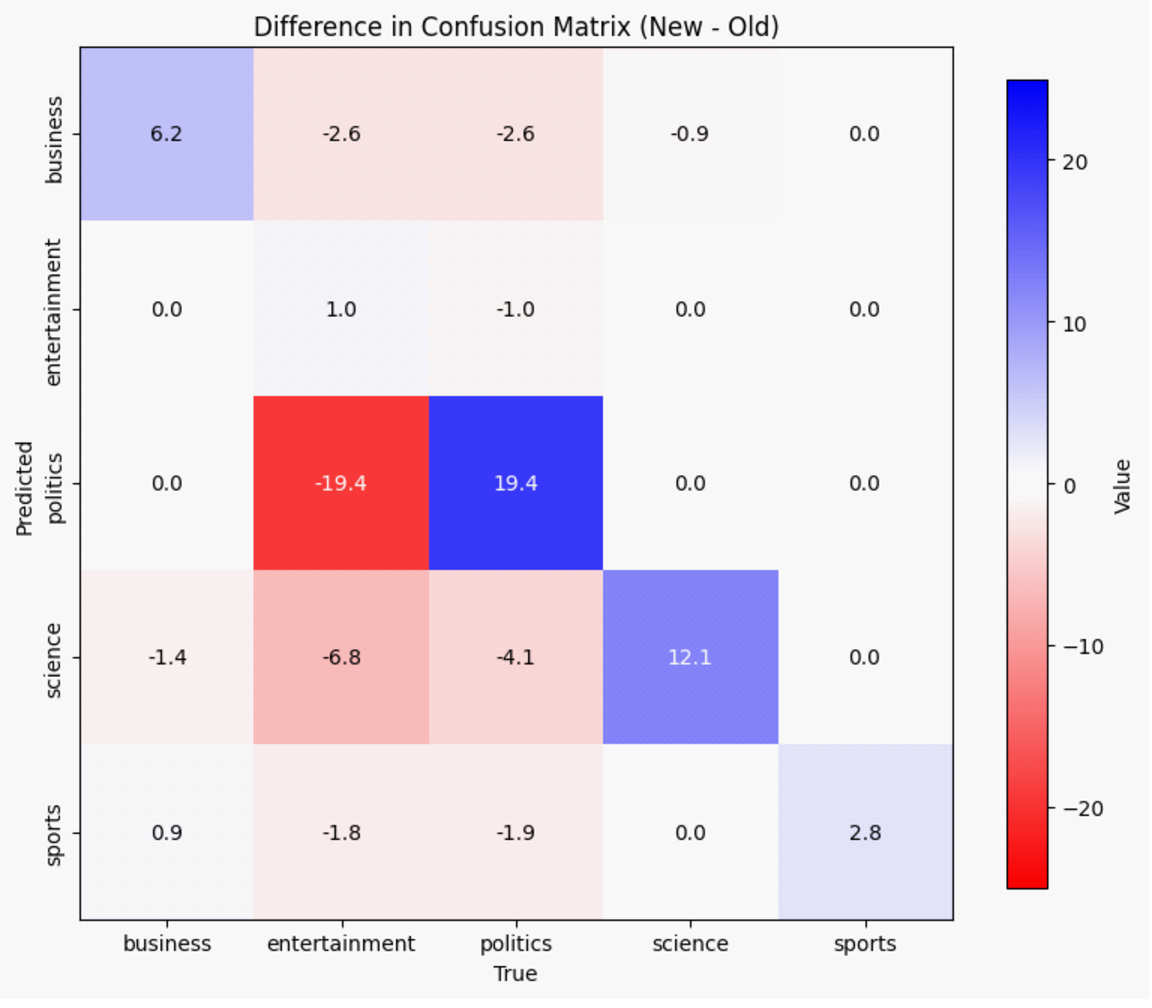}
\caption{Difference in model performance degradation between our method and an existing debiasing method.}
\label{p4:06}
\end{figure}

Figure \ref{p4:06} shows the confusion matrix of the debias results from our method minus the confusion matrix of the debias results from Hard Debias. The vertical axis plots the correct answer labels while the horizontal axis plots the predicted labels.
We found that categorization accuracy has improved in all categories since the diagonal components are all positive.

\begin{table}[hbtp]
  \caption{Comparison of Accuracy and F1 Score of Hard Debias with our method.}
  \label{table:4_03}
  \centering
  \begin{tabular}{lcr}
    \hline
      & Accuracy  &  F1 Score  \\
      &&(weighted)\\
    \hline \hline
    Our method  & 0.93252  & 0.93254 \\
    Hard Debias  & 0.85481   & 0.85743 \\
    \hline
  \end{tabular}
\end{table}

Table \ref{table:4_03} compares the Accuracy and F1 Score of the models debiased by our method with those debiased by the Hard Debias method. The Accuracy and F1 Score increased almost identically for the model with our method, indicating that this method equally suppressed model performance degradation for all categories.

\subsection{Discussion}
\label{sec_4_5}
\noindent

This paper reported the results of classifying bias in three different ways in Section \ref{sec_4_3}. First, we performed bias classification by the k-means method, but the optimal number of clusters could not be determined.  After this trial, we adopted a word categorization method using the third categorization model.
Then, as described in Section \ref{sec_4_4}, we found that the analysis of the complete debiasing results shows that the degree of influence of debiasing varies from category to category, and that the performance of the model is also degraded.  On the other hand, we could achieve debias with minimal degradation of model performance by adjusting the degree of debias for each category based on the visualization results. Furthermore, we found that our method can flexibly respond to patterns in which debiasing is emphasized, performance degradation is minimized, or both are equally emphasized, by visualizing and comparing the degree of bias relaxation and model performance degradation for each degree of debiasing. 

Our experiment showed that the effect of debiasing on the ``entertainment'' category was small.
The reason of this result seems that the number of words belonging to the ``entertainment'' category is large in the dataset used in our study, and even if some information is lost due to debiasing, the word group as a whole is unlikely to be far from the semantic space of ``entertainment''. The ambiguity of the definition of ``entertainment'' could also be a factor, so it is possible that the effect of debiasing could be analyzed more precisely by setting the category in a more subdivided form. On the other hand, the fact that the business category with the smallest number of data showed higher accuracy after debiasing than science and politics indicates that we cannot simply say that the effect of debiasing becomes smaller as the number of data increases. In other words, it is necessary to discuss not only the number of words but also other factors such as how the category itself is defined.

\section{Conclusion and Future Work}

This paper presented a visualization-supported debiasing method that allows the degree of debiasing to be adjusted for each feature of embedded words and its parameters of a word categorization task. Compared to conventional debiasing methods, this method has flexibility in selecting the degree of debiasing for each word feature while minimizing model performance degradation.  This paper demonstrating the usefulness of the proposed debiasing method and the effectiveness of the visualizations.

As a future issue, we would like to adopt further category classification methods. In particular, further study is needed on visualization methods and specific category classification methods when the number of categories increases.

Another future work is development of a user interface that allows the user to adjust the degree of debiasing with an interactive mechanism while observing the visualizations that represent the classification results.
It was difficult to archive a practical response time of the debiasing process and therefore we have not develop such an interactive system.  We tried to speed up the process by using faster algorithms, but this resulted in a loss of accuracy.  In the future, we would like to develop a system that allows the user to freely adjust the degree of debiasing for each word feature while observing the visualizations, and achieve both accuracy and speed on the debiasing process.

\end{document}